\documentclass[12pt, onecolumn]{IEEEtran}

\usepackage{graphicx}
\usepackage{amsmath}
\usepackage{amsmath, amsfonts,epsfig, multirow, floatflt}
\usepackage{epsfig,graphics,graphicx,latexsym,amsfonts,amssymb,amsmath,verbatim,soul}
\usepackage{subfigure}
\usepackage{cite}
\usepackage{makecell}

\graphicspath{{figures/}}

\begin{document}
\baselineskip 24pt
\parskip 9pt
\thispagestyle{empty}

%%**************************
\newpage

\baselineskip 19pt
\parskip 4pt

\setcounter{page}{1}

\begin{center}
\vspace*{0mm}

{\LARGE \bf  Reliable and Efficient Autonomous Driving: the Need for Heterogeneous Vehicular Networks} \vspace*{10mm}

{\normalsize Kan Zheng$^\ast$, {\it Senior Member, IEEE} Qiang Zheng$^\ast$, Haojun Yang$^\ast$, Long Zhao$^\ast$, Lu Hou$^\ast$, and Periklis Chatzimisios {\it Senior Member,~IEEE}$^\ddag$\\

\vspace{0.3cm}

\small $^\ast$Wireless Signal Processing and Network Lab\\
Key laboratory of Universal Wireless Communication, Ministry of Education \\
Beijing University of Posts \& Telecommunications\\
Beijing, China, 100088 \\

$^\ddag$Department of Informatics\\
Alexander Technological Educational Institute of Thessaloniki (ATEITHE)\\
P.O. Box. 141 , 57400 Sindos\\
Thessaloniki, Greece\\

E-mail: \tt{kzheng@ieee.org} \\
}
\end{center}

\vspace*{10mm}

\begin{center} {\bf Abstract}

\end{center}

Autonomous driving technology has been regarded as a promising solution to reduce road accidents and traffic congestion, as well as to optimize the usage of fuel and lane. Reliable and high efficient Vehicle-to-Vehicle (V2V) and Vehicle-to-Infrastructure (V2I) communications are essential to let commercial autonomous driving vehicles be on the road before 2020. The current paper firstly presents the concept of Heterogeneous Vehicular NETworks (HetVNETs) for autonomous driving, in which an improved protocol stack is proposed to satisfy the communication requirements of not only safety but also non-safety services. We then consider and study in detail several typical scenarios for autonomous driving. In order to tackle the potential challenges raised by the autonomous driving vehicles in HetVNETs, new techniques from transmission to networking are proposed as potential solutions.

\begin{flushleft}
\end{flushleft}
\newpage
\baselineskip 24pt
\parskip 9pt

\clearpage
\section{Introduction}

The automotive industry has recently shifted from developing advanced vehicles to safe and comfortable ones, which stimulates the development of new intelligent vehicles with autonomous driving control~\cite{citation_1}. The Autonomous Driving Vehicle (ADV) is a multidisciplinary product that can integrate  automotive control, information processing, communication capabilities and so on. Governments and society can be substantially benefited from autonomous driving, including prevention of road accidents, reduction of traffic congestion, as well as optimal usage of fuel and lane~\cite{citation_2}. In order to realize autonomous driving, vehicles need to be capable of sensing the surrounding environment as well as performing control and path planning without any human intervention~\cite{citation_3}. Global automakers and information technology companies, such as General Motors, Volkswagen, Toyota and Google, expect to have ADVs on the market in 2020 and 25\% of the vehicles out on the road to be ADVs by 2035~\cite{citation_4}, which coincides with the timetable of the Fifth Generation (5G) wireless communication systems. \par

Nevertheless, several challenges still need to be conquered for autonomous driving~\cite{citation_5}, such as: 1) to have knowledge of the exact position of the vehicle and to decide how to reach the destination ; 2) to sense the surrounding environment in order to avoid a vehicle collision; 3) to detect the road signs as well as lanes, crosswalks, speed bumps and so on. Currently, in order to face these challenges, sensor systems with cameras, radar or laser range finder, and advanced autonomous driving algorithms are employed. However, it is still far from enough since the driving behavior of vehicles is significantly affected by the surrounding vehicles and this is not well exploited due to the limited communication ability between vehicles. Moreover, the main approach to detect the surrounding environments is performed by utilizing sensor systems but is highly limited by the environment in which vehicles operate, e.g., obstacles, other vehicles, weather conditions and so on.\par

Thanks to the rapid development of wireless communication technologies, vehicular networks are expected to boost the development of autonomous driving and employ the Vehicle-to-Vehicle (V2V) and Vehicle-to-Infrastructure (V2I) communication techniques, which can be effectively used to detect surrounding conditions. The autonomous driving vehicle may become safer if it makes autonomous decisions with reliable information provided by vehicular networks. For example, every vehicle can periodically broadcast safety-related messages about its current condition to its neighboring vehicles, which is helpful for all vehicles in order to accurately know their surrounding environment. On the other hand, vehicular networks can significantly improve traffic efficiency. Moreover, passengers in autonomous driving vehicles are likely to enjoy infotainment content during their journey by accessing Internet mainly via the use of V2I communication.\par

Due to high mobility and the dynamic change of the network topology, it is difficult to provide satisfactory services only through a single wireless access network, such as the Dedicated Short Range Communication (DSRC) or the the Third Generation (3G) Long Term Evolution (LTE) networks~\cite{dsrc}\cite{citation_7}\cite{d2d}. In particular, 1) DSRC networks are mainly designed for short-range communication without considering the pervasive communication infrastructure; 2) The delay incurred in LTE networks may be significantly deteriorated with a large number of vehicles while the strict latency for delivering real-time information for the autonomous driving is required; 3) The huge volume of data generated by sensors in ADVs is beyond the capacity of the current vehicular networks.\par

Therefore, this paper presents Heterogeneous Vehicular NETworks (HetVNETs), which integrate different types of wireless access networks, such as LTE and DSRC, in order to satisfy the various communication requirements of autonomous driving~\cite{citation_8}\cite{hetvnet}. We also improve the existing protocol stacks and define certain types of messages in HetVNETs that are essential to support the autonomous driving. Taking advantage of these different message types, the ADVs can achieve the desired traffic behavior such as overtaking, changing lane and so on. Moreover, several typical autonomous driving scenarios are discussed and thoroughly analyzed through studying their specific traffic and communication characteristics. Moreover, in order to ensure reliable low-latency message delivery and efficient provision of high data rate transmission for ADVs, several advanced techniques are correspondingly proposed for HetVNETs.\par

\section{HETVNET for Autonomous Driving}

\subsection{Network Infrastructure}

\begin{figure}[h]
	\centering
	\includegraphics[width=0.7\textwidth, angle=0]{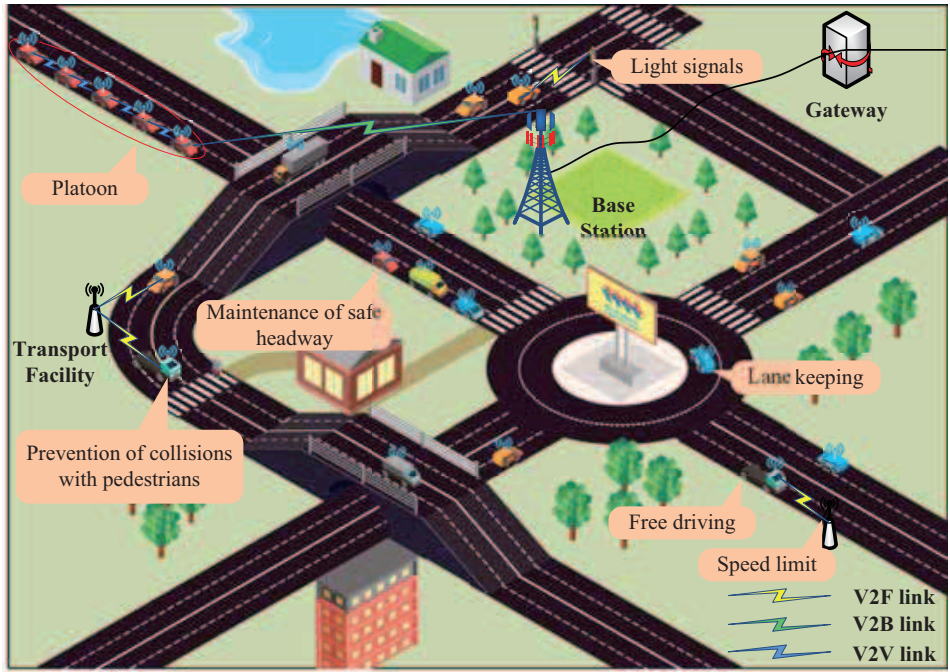}
	\caption{Illustration on HetVNET infrastructure for the autonomous driving.}
	\label{fig_1}
\end{figure}

As shown in Fig.~\ref{fig_1}, there are mainly three types of vehicles in HetVNETs for autonomous driving, i.e., \par

\subsubsection{Manually Driving Vehicle with communication modules (MDV)}
A MDV is fully controlled by humans. Different from the traditional vehicles, each MDV is equipped with an appropriate communication module, e.g., having both DSRC and LTE communication, which support real-time information exchange among neighboring vehicles as well as between vehicles and Base Stations (BSs).  \par

\subsubsection{Autonomous Driving Vehicle (ADV)}
An ADV has the ability to cruise safely without human intervention. Apart from the communication module, another five basic modules are usually needed to support autonomous driving, i.e., perception, localization, planning, control and system management~\cite{citation_1}. Perception is the process that obtains a clear view of the surrounding environment via various techniques such as radar, lidar and vision detection. Localization can be implemented by using a Global Positioning System (GPS). Based on the information obtained from perception and localization modules, the navigation behavior of an ADV can be determined via a planning module. The control module executes the desired command received from the planning module; and the system management module supervises the overall state of the autonomous driving system. \par

\subsubsection{Platoon}
A platoon consists of a group of vehicles, i.e., one head vehicle and several followers. The head vehicle can be either a MDV or an ADV, while the followers incorporate automatic longitudinal speed control and their lateral movements are controlled by the header. In order to safely control the followers, the head vehicle can interrupt the platoon mode at any time. In addition, the safe headaway between vehicles must be maintained as well as other safety requirements. The required communication capabilities of a platoon are to broadcast its active state using safety messages, to receive these messages, and to establish unicast sessions.  \par

To achieve autonomous driving, it is very important to let the ADVs reliably identify the behavior of other vehicles. Hence, the communication for information exchange is critical, which may happen in the following different ways, i.e.,\par

\setcounter{subsubsection}{0}
\subsubsection{V2V communication}
It provides an efficient way for ADVs to share information between each other, which can help to enhance safety, reduce traffic congestion and avoid vehicle collisions. \par

\subsubsection{V2I communication}
According to different types of network infrastructure, V2I communication can be further divided into V2F for local communication and V2B for global communication as follows, i.e.,\par

\begin{itemize}
	\item{\emph{Vehicle-to-Facility (V2F):}}
	The communication module can also be installed at road facilities (e.g., speed limit signs, and traffic lights), which provide the capability to disseminate periodical and event-driven messages. The main functionality of V2F communications is to broadcast warning messages to specific road sections, speed limit notification and traffic light signals to nearby vehicles in order to avoid accidents. Another functionality is to act as relay link to improve the communication reliability in the given areas such as intersections with high buildings or obstacles. Also, V2F can be used by transport management Departments to regulate the vehicles on the road.
	\item{\emph{Vehicle-to-Base Station (V2B):}}
	V2B communication mainly refers to the wireless link between vehicles and BSs. Through V2B links, the vehicles can access the Core Network (CN) and Internet. It is mainly supported by cellular networks and plays an important role for autonomous driving. Most of non-safety related services are provided through V2B links. Based on the information obtained via V2B links, the service center can obtain a global view of the traffic network and give useful insights to vehicles such as optimal navigation paths.

\end{itemize}

By integrating HetVNETs into the autonomous driving system, the travel efficiency and safety of the vehicles is envisioned to be significantly improved.\par

\subsection{Types of Autonomous Driving Messages}

\begin{figure}[h]
	\centering
	\includegraphics[width=0.6\textwidth, angle=0]{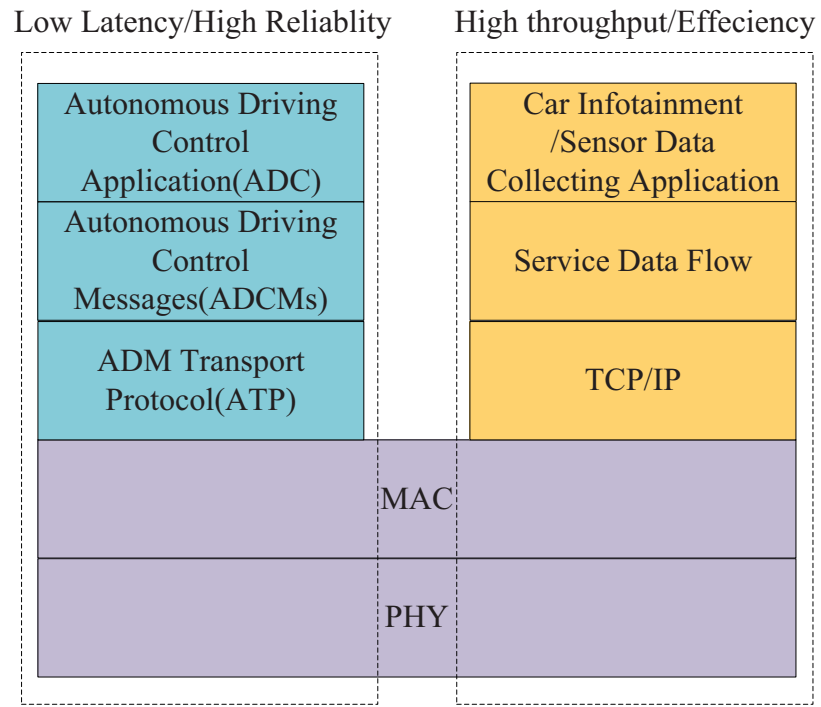}
	\caption{Protocol stack to support ADMs in HetVNETs.}
	\label{fig_2}
\end{figure}

Fig.~\ref{fig_2} presents the proposed HetVNETs protocol stack for vehicular networks, which is expected to support the various types of messages for autonomous driving. Autonomous Driving Control (ADC) applications are in charge of the control and management of ADVs. The layer of Autonomous Driving Control Messages (ADCMs) is used in order to support these applications. Similar to Wireless Access in Vehicular Environments (WAVE) Short Message Protocol (WSMP) in DSRC~\cite{WSMP}, efficient ADCM Transport Protocol (ATP) is designed since it is critical to deliver ADCMs with low latency for high mobility scenarios. One of the basic functions of ATP is to provide the broadcasting services without connection establishment, which facilitates the message dispensed between ADVs or from ADVs to network infrastructure. On the other hand, the passengers of ADVs want to use the entertainment services during traveling. Meanwhile, ADVs are equipped with a large number of sensors resulting in the generation of massive volumes of sensor data. Thus, there is a great demand for HetVNET to support the applications that have high throughput and efficiency requirements. \par

As shown in Table. \ref{table_1}, ADCMs can be roughly categorized into two types, i.e., \par
\begin{itemize}
	\item{\emph{Periodic State Messages (PSMs):}}
	PSMs are mainly employed to indicate the state information of vehicles, such as position and traveling directions. This information can be collected by the neighboring ADVs in order to estimate safety factors before making any action and it is also broadcasted to infrastructure via V2B links. Based on the PSMs, the service center can make data analysis, gather statistics of traffic flow and so on.
	\item{\emph{Action-Triggered Messages (ATMs):}}
	ATMs include the action contents of the ADVs, which can be used for decision in the next moment. Only by utilizing these messages, an ADV can accurately know the surroundings as well as the movement of other ADVs. Thus, it can make the appropriate reaction autonomously and then send its changing state to other vehicles.
\end{itemize}

For the sake of the illustration, a few examples are also given in Table. \ref{table_1}. \par

\section{Typical Scenarios of Autonomous Driving}
In order to understand the requirements of ADVs in HetVNETs, it is necessary to study several typical application scenarios with autonomous driving. Thus, three scenarios, namely highway free flow, highway synchronized flow and urban intersection, are discussed in this Section, where both traffic and communication characters are analyzed. \par

\subsection{Scenario 1: Highway Free Flow}

\subsubsection{Traffic Characters}
\begin{figure}
\centering
\subfigure[Scenario 1: Highway Free Flow.]{\includegraphics[width=0.35\textwidth, angle=0]{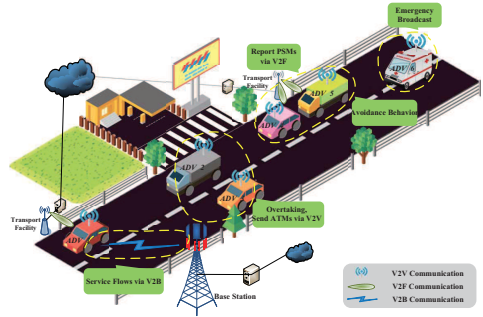}
\label{fig_Scenario1} }\\
\subfigure[Scenario 2: Highway Synchronized Flow.]{\includegraphics[width=0.35\textwidth, angle=0]{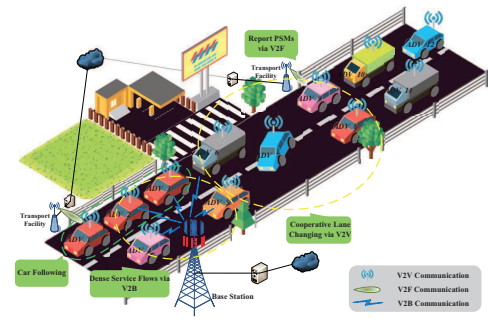}\label{fig_Scenario2}}\\
\subfigure[Scenario 3: Urban Intersection.]{\includegraphics[width=0.35\textwidth, angle=0]{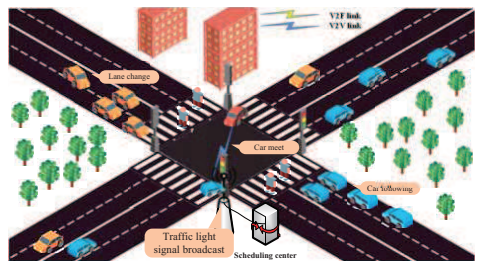}\label{fig_Scenario3}}
 \caption{Illustration on typical scenarios of autonomous driving vehicles on the road.} \label{fig_Scenario}
\end{figure}

As illustrated in Fig.~\ref{fig_Scenario1}, the number (density) of ADVs is very small (low) in a highway free flow scenario. In general, four types of ADVs traffic behaviors exist, i.e.,\par

\begin{itemize}
	\item{\emph{Normal Driving Behavior:}}
	ADVs can travel freely with the desired speed, which is not constrained by other vehicles on the road. This is the main behavior in the free flow scenario.
	\item{\emph{Overtaking Behavior:}}
	In order to maintain the desired speed, ADVs sometimes need to overtake other vehicles, which travel with relative low speed. There are two phases in the overtaking action. The first phase is lane changing with safety when the adjacent lanes are vacant. The second one is to accelerate and surpass the heading vehicle until it acquires a safe distance and then changes back to the original lane. As illustrated in Fig.~\ref{fig_Scenario1}, ADV 3\# is driving behind ADV 2\#, whose speed is lower than the one of ADV 3\#. In this case, ADV 3\# may overtake ADV 2\# for sake of maintaining its desired speed. Therefore, ADV 3\# needs to firstly inform ADV 2\# about its action through V2V communication, and then checks whether the adjacent passing lane is safe or not. If the adjacent passing lane is clear, it can execute lane-change action to overtake another ADV. Otherwise, it must wait for a while before the next attempt.
	\item{\emph{Avoidance Behavior:}} For the purpose of safe driving, avoidance behavior is essential as a driving behavior of the highway free flow. For example, when ADV 4\# and ADV 5\# are too close in Fig.~\ref{fig_Scenario1}, ADV 5\# has to slow down to avoid bumping ADV 4\#.
	\item{\emph{Emergency  Avoidance Behavior:}} When ADVs are travelling, they may encounter some emergency vehicles, e.g., ADV 6\# in Fig.~\ref{fig_Scenario1}. At this moment, ADVs have to reduce their speed and pull over immediately in order to ensure that emergency vehicles quickly pass by. Thus, ADV 6\# broadcasts the appropriate emergency messages to other vehicles all the time when it is moving. This behavior may happen in all three scenarios and, thus, it is not discussed in the next scenarios.
\end{itemize}

\subsubsection{Communication Characters}

Each traffic behavior has its specific communication requirements in order to achieve the corresponding action safely. Therefore, different communication ways with the corresponding message types need to be applied to guarantee the success of each traffic behavior. For example, an ADV with normal driving behavior needs to distribute its PSMs through both V2V and V2I links. Moreover, it is necessary for an ADV to send ATMs via V2V communication to ensure safety when it performs the overtaking. \par

Moreover, services such as entertainment applications can be enjoyed mainly by the occupants through V2B communication. However, due to the high speed of the vehicles in the free flow, the fast fading propagation effects of the radio channels are quite serious, which significantly deteriorate the quality of communication links. Therefore, it becomes a challenge how to guarantee reliable communication under such scenario.

\subsection{Scenario 2: Highway Synchronized Flow}

\setcounter{subsubsection}{0}
\subsubsection{Traffic Characters}
 As illustrated in Fig.~\ref{fig_Scenario2}, the number of vehicles in the highway synchronized flow scenario is larger than that in the free flow, which results in higher density and lower speed. Generally, the synchronized flow contains two traffic characters: a) vehicles are in flux while their speed is relative high; b) the speed of all vehicles in different lanes tends to be synchronized. Therefore, ADVs travelling in this scenario have very limited flexibility. Such characters give rise to three types of typical traffic behaviors, i.e., \par
\begin{itemize}
	\item{\emph{Car Following Behavior:}}
	Due to the converging speed, ADVs have to follow the front vehicles, which is the most common behavior under this scenario. In Fig.~\ref{fig_Scenario2}, there are a few ADVs with car following behavior, e.g., ADV 1\#, ADV 2\# and ADV 3\#. If ADV 1\# encounters an unexpected event and thus has to slow down, ADV 2\# and ADV 3\# have to also reduce their speed correspondingly in order to keep the safe headway.
	\item{\emph{Lane Changing Behavior:}}
	In the synchronized flow, overtaking of ADVs is very difficult due to the high vehicle density. However, in order to improve traffic efficiency, some ADVs with relatively high speed are likely to change lane, e.g., ADV 7\# in Fig.~\ref{fig_Scenario2}.
	\item{\emph{Avoidance Behavior:}} This behaviour is very important to avoid car crash especially in this scenario. Different from the free flow scenario, an ADV, e.g., ADV 8\# in Fig.~\ref{fig_Scenario2}, is likely to encounter a collision possibly due to lane changing as well as the car following. \par
\end{itemize}

\subsubsection{Communication Characters}
 In order to guarantee traffic safety and efficiency, cooperative communication among vehicles is likely to happen, which is quite different from the highway free flow scenario. For instance, cooperative lane changing can decrease the time of direct lane changing to maximize traffic efficiency. Compared to the free flow scenario, besides the severe fast fading prorogation effects, the high density of vehicles makes network topology more complex. Meanwhile, a huge amount of communication requests are generated by the different ADVs on the road. All of these make the target of reliable and efficient communications extremely difficult. Novel communication techniques are desired to be applied especially in this scenario.\par

\subsection{Scenario 3: Urban Intersection}

\setcounter{subsubsection}{0}
\subsubsection{Traffic Characters}
As illustrated in Fig. 3(c), at the intersection ADVs not only interact with other vehicles, but also with the transport facilities and pedestrians. Thus, the situation becomes much more complicated compared with those in highway scenarios. The behavior of vehicles and pedestrians is regulated by traffic lights. The driving speed (e.g., the maximum speed is 40 $km/h$) is much lower than the one in highway scenarios. Moreover, the overtaking is seldom performed due to the high traffic load at the intersection. The following behaviors commonly happen, i.e., \par
\begin{itemize}
	\item{\emph{Car Following Behavior:}}
	This behavior usually occurs when vehicles are queuing in the lane to take a turn. The key message before executing this behavior is related with the accurate position of the car ahead.
	\item{\emph{Lane Changing Behavior:}}
	When a vehicle drives on the lane that is not its target lane, it has to change lane. For example, a vehicle has to go to the left lane if it wants to turn left.
	\item{\emph{Car Meeting Behavior:}}
	The car meeting behavior is complicated, since two vehicles are involved in and interact with each other. For example, turning right has the higher priority than turning left when the target lanes are the same.
\end{itemize}

\subsubsection{Communication Characters}
The stop-and-go control at an intersection ensures the safe crossing of the intersection. However, it may introduce the inconvenience of frequent stops and idling until achieving the right-of-way, which significant reduces traffic efficiency. Through efficient communication via either V2V or V2I links in an intersection, each vehicle may have adequate maneuver commands in real time. Therefore, traffic operations at an intersection without stop-and-go-style traffic lights and signs become available when the road is full of ADVs.\par

\section{Potential Challenges and Solutions}
HetVNETs utilizing improved protocols and messages provide the feasibility to support the behavior of ADVs. However, new communication challenges arise when the Quality of Service (QoS) requirements of both ADC messages and other services need to be supported in various scenarios. To tackle these challenges in HetVNETs, novel techniques from the signal transmission to networking, are then discussed as possible solutions.\par

\subsection{Low-Latency and High-Reliable Transmission Techniques}

In autonomous driving systems, different types of messages or information have different QoS requirements, such as low latency and high reliability for safety messages, high data rate for non-safety multi-media applications. It is hard to meet all these requirements only by the existed transmission techniques in either LTE or DSRC networks, which motivates the development of new transmission techniques in HetVNET for implementing autonomous driving. \par

Filtered-Orthogonal Frequency Division Multiplexing (F-OFDM) applies sub-band digital filter to shape the spectrum of sub-band OFDM signal, which has the good out-of-band leakage rejection and thus supports asynchronous OFDM Access (OFDMA) transmission without timing advanced signal \cite{citation_10}. On the other hand, through joint optimization of multi-dimension Quadrature Amplitude Modulation (QAM) and non-orthogonal sparse codewords, Spares Code Multiple Access (SCMA) is capable of multiplexing more users and improving system reliability \cite{citation_11}. Therefore, the combination of SCMA and F-OFDM is expected to be one of the possible candidate techniques to satisfy the communication requirements of HetVNETs in autonomous driving.\par

\begin{figure}
\centering
\subfigure[Transmitter.]{\includegraphics[width=0.4\textwidth, angle=0]{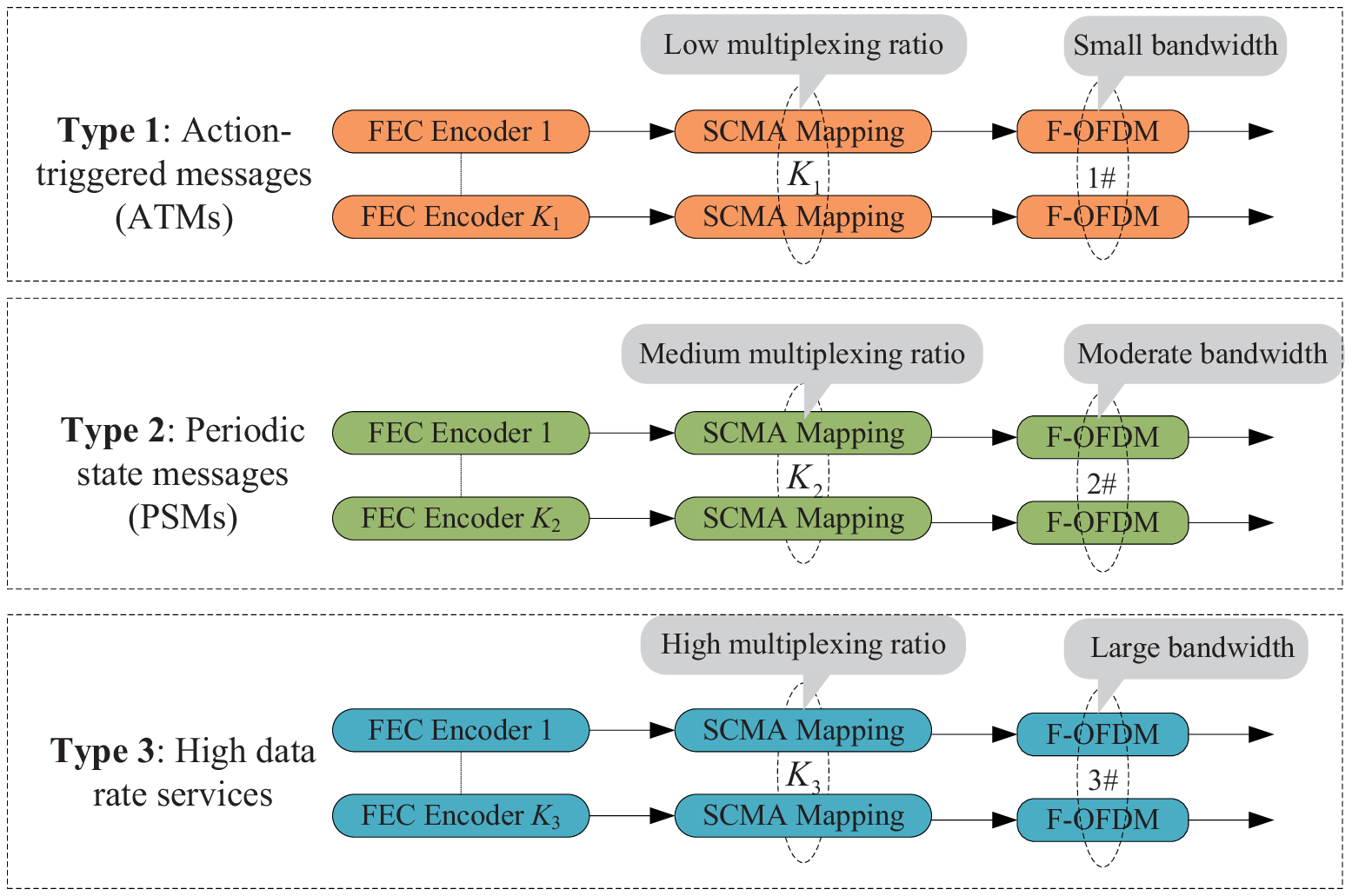}
\label{fig4a} }
\subfigure[Bandwidth.]{\includegraphics[width=0.3\textwidth, angle=0]{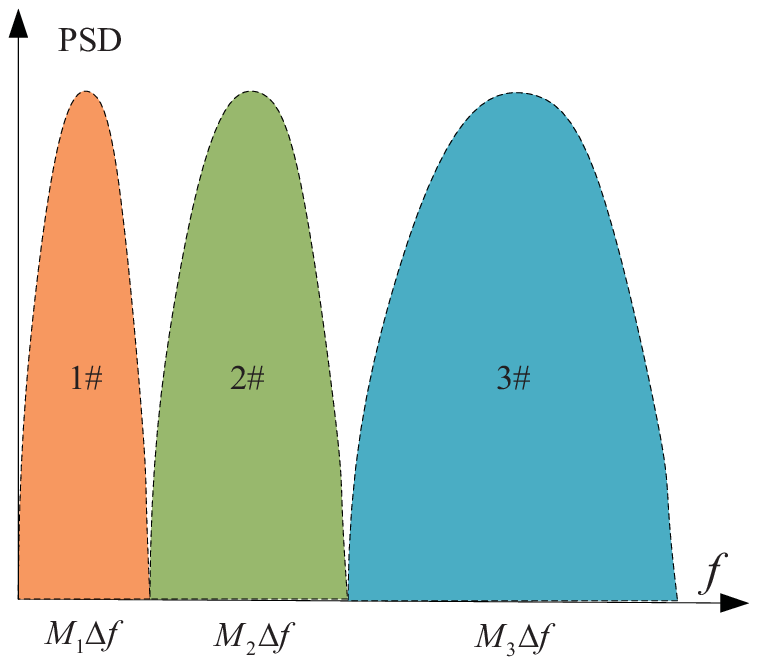}\label{fig4b}}
 \caption{A paradigm of transmission design for HetVNETs.} \label{fig4}
\end{figure}

Fig. \ref{fig4a} presents an example of the transmitter based on the combination of SCMA and F-OFDM, which supports three types of services. The three corresponding subbands are also shown in Fig. \ref{fig4b}, which are called as \textbf{Subband 1\#} with bandwidth ${M_1}\Delta f$, \textbf{Subband 2\#} with bandwidth ${M_2}\Delta f$, and \textbf{Subband 3\#} with bandwidth ${M_3}\Delta f$ (where $\Delta f$ is the subcarrier bandwidth and ${M_1} \le {M_2} \le {M_3}$). The ATMs and PSMs with low latency and high reliability can be transmitted in \textbf{Subband 1\#} and \textbf{Subband 2\#}, respectively. Meanwhile, the vehicles with non-safety multimedia applications of high date rate are served in \textbf{Subband 3\#}. Different number of vehicles, i.e., ${K_1},\ {K_2},\ {K_3}$, may be scheduled in different subbands. Usually the number of multiplexing vehicles in \textbf{Subband 1\#} and \textbf{Subband 2\#} are chosen to be small while no such limitation exists in \textbf{Subband 3\#}. \par

This paradigm is capable of satisfying low-latency, high-reliability and massive access for autonomous driving. Firstly, the asynchronism character of F-OFDM and the grant-free access of SCMA, i.e., the message or information transmission without signaling overhead between vehicles, such as grant request, and acknowledgement, can significantly reduce delay of messages delivery. Moreover, a small number of vehicles in a single subband may decrease the collision probability of SCMA codewords, which can improve the decoder performance of Message Passing Algorithm (MPA) and, thus, increase transmission reliability. Moreover, F-OFDM is capable of utilizing the fragmental spectrum resources and shaping flexible bandwidth for different kinds of services, which maximizes spectrum efficiency. Thus, together with SCMA multiplexing more users, F-OFDM transmission is capable of supporting massive access of ADVs. \par

Besides, SCMA and F-OFDM do not change the essential transmission character of either downlink OFDM or the uplink OFDMA. The combination of SCMA and F-OFDM keeps the good backward compatibility to existing LTE systems. Therefore, the SCMA/F-OFDM transmission provides a feasible evolutional roadmap from LTE to HetVNETs. \par

\subsection{Cooperative Autonomous Driving Techniques}

In the complicated vehicular environments, ADVs need to have an in-depth understanding of their surrounding environment to make optimal cooperative driving decisions and path scheduling. However, the intrinsic limitations of traditional information perception such as camera and radar often prevent cooperative decisions. Therefore, in this subsection, a cooperative autonomous driving framework based on HetVNETs is proposed, where each ADV can share information both locally for traffic safety and globally for traffic efficiency. \par

\begin{figure}
\centering
\subfigure[Example scenario.]{\includegraphics[width=0.35\textwidth, angle=0]{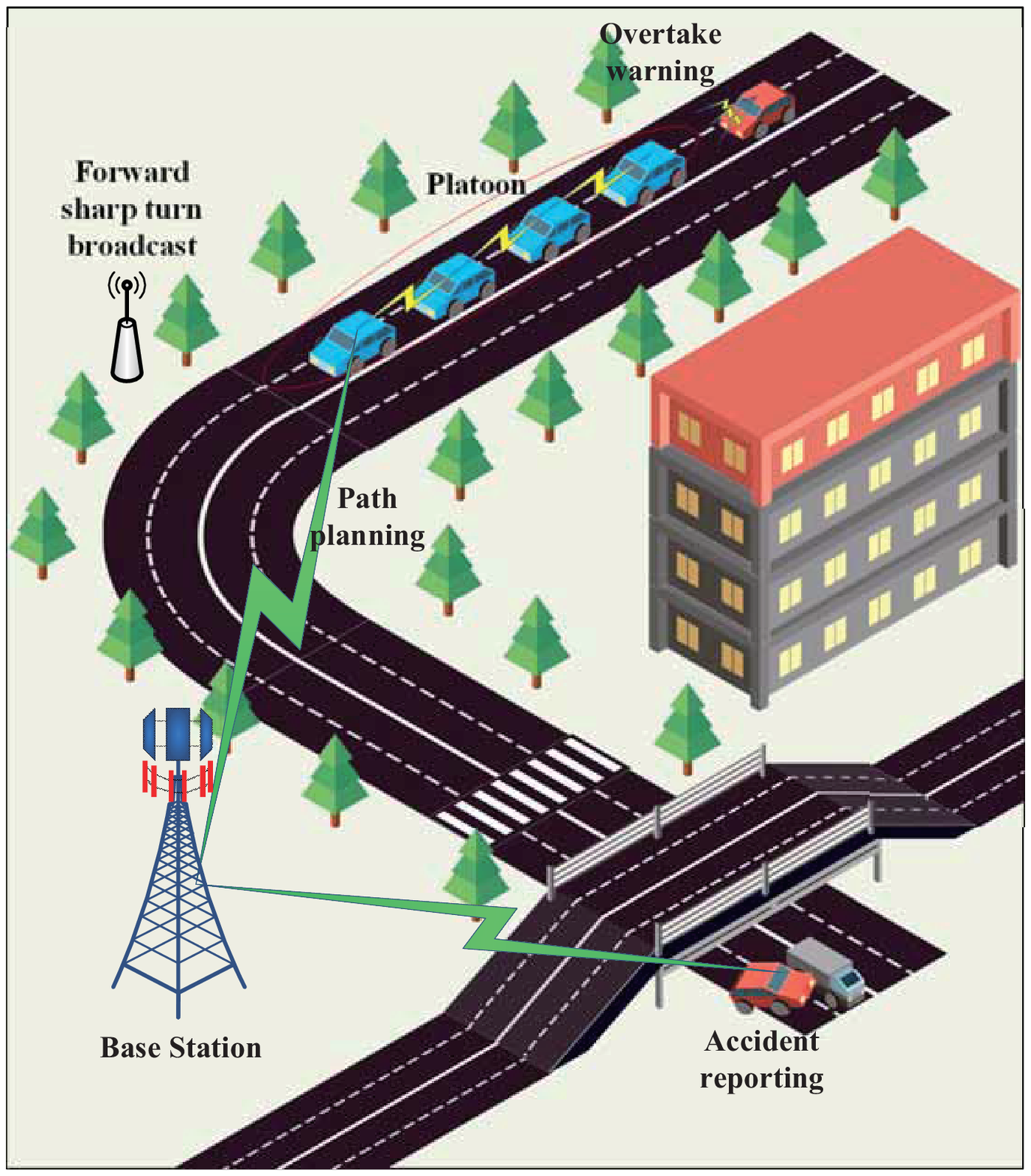}
\label{fig_Coopa} }
\subfigure[Typical functions.]{\includegraphics[width=0.42\textwidth, angle=0]{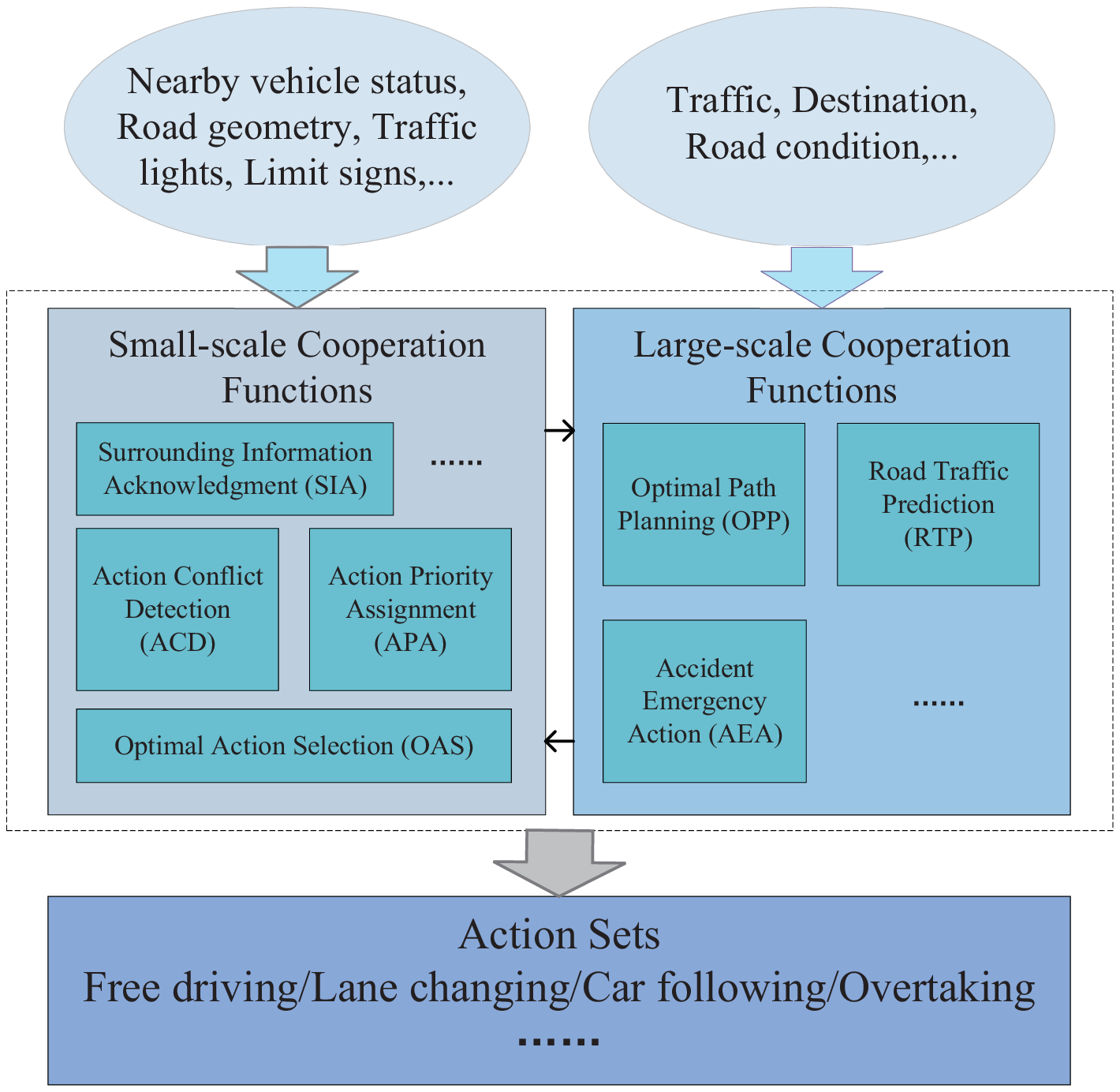}\label{fig_Coopb}}
 \caption{Illustration of hierarchical cooperative autonomous driving scheme.} \label{fig_Coop}
\end{figure}

As illustrated in Fig.~\ref{fig_Coopa}, the proposed hierarchical cooperation can be divided into two layers, i.e., small-scale and large-scale cooperation. The former is executed only within the local area and with the fine time resolution in order to ensure safety. The latter happens across the large geographical area with the coarse time resolution in order to enhance traffic efficiency. \par

\subsubsection{Small-scale cooperation}
The main objectives of small-scale cooperation are to guarantee traffic safety through cooperation between vehicles in the local area. Such cooperation is implemented in a distributed manner, which significantly reduces signal overhead between vehicles. As illustrated in Fig.~\ref{fig_Coopb}, there are several typical functions needed to support small-scale cooperation, i.e.,\par

\begin{itemize}
	\item{\emph{Surrounding Information Acknowledgement (SIA):}}
	SIA can provide accurate environment information for the next step that can be classified into two types, i.e., dynamic and static information. The former includes nearby vehicle state information and event-driven messages, which are transmitted via V2V link. While the latter mainly contains the information such lane, intersection, and speed limit, which is provided by V2I communication.\par
	\item{\emph{Optimal Action Selection (OAS):}}
	It generates an optimal action for the next interval, which may need to solve an optimization problem based on obtained surrounding information and events. The action set mainly includes free driving, lane changing, lane keeping, car following, overtake, platoon and so on.
	\item{\emph{Action Conflict Detection (ACD):}}
	A vehicle checks its required action with those of neighboring vehicles. Only when there is no conflict, the control command for this action can be executed.
	\item{\emph{Action Priority Assignment(APA):}}
	When the conflict happens between the required actions of the vehicles, only the action with the higher priority may be chosen.
\end{itemize}

\subsubsection{Large-scale cooperation}

It aims to disseminate the information over a large geographical area to improve traffic efficiency. Furthermore, a few of functionalities such as path prediction and scheduling capabilities of involved vehicles can be leveraged when the upcoming traffic congestion can be detected in advanced via large-scale cooperation. Different from small-scale cooperation, it is executed in a centralized manner via V2B links. The framework of the large-scale cooperation is illustrated in Fig.~\ref{fig_Coopb}. Firstly, the cloud server collects information such as road conditions, unexpected traffic congestion, adverse weather conditions, traffic density, via V2B links. Then, it calculates the corresponding results for different applications. There are a few of functions to support the larger-scale cooperation, e.g., Optimal Path Planning (OPP), Road Traffic Prediction (RTN), and Accident Emergency Action (AEA). \par

%The platoon may involve both large-scale and small-scale cooperation. In the small-scale cooperation, the vehicles in a platoon %transmit and receive periodical PSMs from each other and keep the safe headway. In the large-scale cooperation, the header of a %platoon can receive messages from the BSs, including path planning and road condition for improving traffic efficiency on the road. %\par

\subsection{Layered-Cloud Computing Techniques}
It is estimated that an ADV generates around 1 Gigabyte data per second (Gbps), which mainly comes from sensors. To store such amount of data in the vehicle during traveling needs extremely a huge local storage unit. Therefore, the Remote Cloud (RC) is proposed as a feasible solution by the aid of the offloading techniques over high throughput wireless transmission. In particular, it provides abundant communication and computing resources in order to ensure the safety and traffic efficiency of ADVs. However, when ADVs become more and more popular, thousands of ADVs may be present on the road and simultaneously generate sensor data. Thus, it is impractical to transmit all the sensor information of each ADV over V2I links, which is a very important challenge even for 5G networks with the peak rate of 10 Gbps. Therefore, the wireless links between ADVs and the RC have to be efficiently utilized. \par

On the one hand, since the data generated by ADVs has the substantial correlation in the time domain, it is possible to process and compress data before transmission over V2I links. For example, when the sensor data change continuously in time, the ones with very small variation can be omitted. On the other hand, another obvious feature of the generated data is its local interests, which means that only ADVs in the vicinity are likely to enjoy the common interests such as local traffic congestion and road condition messages. Therefore, the data of common interests can be kept local rather than been uploaded to the RC, which may greatly reduce the capacity requirements of the V2I links.\par

Moreover, the collaboration in the sharing and processing of sensor data between the ADVs can significantly improve the location accuracy and safety of the driving. Vehicular Cloud Computing (VCC) is a new promising technology to take advantage of cloud computing to serve the vehicles~\cite{citation_13}. The computing and storage resources in VCC can be utilized to enhance the abilities of ADVs. In other words, Vehicular Clouds (VCs) can provide a good platform for the coordinated deployment of the sensor aggregation, fusion and database sharing applications required by ADVs. For example, ADVs can enlarge the sensing coverage by aggregating the data from geographically distributed ADVs.\par

Therefore, a layered-cloud computing architecture for ADVs can be deployed as one of the feasible solutions. It includes not only a RC but also the VCs. The ADVs can send the request of either driving or entertainment to any layer of the cloud. \par

%The following example is given only for the sake of illustration, i.e.,
%\par
%
%\begin{itemize}
%	\item{\emph{Step 1:}}
%	The ADV sends a request to the cloud for the resources to process the desired data.
%	\item{\emph{Step 2:}}
%	 The cloud analyzes how to make the response according to the global status of all the resources. For example, the cloud transfers the request to RC if the resources in local VC are occupied completely, or allocates other ADVs in VC to serve the request if there is some resources available.
%	\item{\emph{Step 3:}}
%	When the ADV receives the requested data successfully, it sends the acknowledge message to the cloud. Then, the cloud update the global state, e.g., the information of which contents is saved in this ADV.
%\end{itemize}

\section{Conclusion}
Reliable and efficient communication is extremely important to guarantee safety and comfortability of ADVs. Therefore, in this paper, we proposed HetVNETs together with an improved protocol stack and new messages that can support autonomous driving. Then, three typical scenarios for the ADVs including highway and urban intersection were discussed. Their specific traffic behavior puts forward the various communication requirements, which raise various technical challenges for HetVNETs. Thus, in order to combat these challenges, new techniques such as F-OFDM/SCMA transmission, hierarchical cooperative driving and layered-cloud computing have been presented for the development of HetVNETs towards the 5G era. \par

\section*{Acknowledgment}
This work is funded in part by National Science Foundation of China (No.61331009), the National High-Tech R\&D Program (863 Program 2015AA01A705), National Key Technology R\&D Program of China (No.2015ZX03002009-004) and Fundamental Research Funds for the Central Universities (No.2014ZD03-02).

\section*{Biography}

\begin{small}

\textbf{Kan Zheng}
(SM'09) is currently a full professor in Beijing University of Posts \& Telecommunications (BUPT), China. He received the B.S., M.S. and Ph.D degree from BUPT, China, in 1996, 2000 and 2005, respectively. He has rich experiences on the research and standardization of the new emerging technologies. He is the author of more than 200 journal articles and conference papers in the field of wireless networks, M2M networks, VANET and so on. He holds editorial board positions for several journals. He has organized several special issues in famous journals including IEEE Communications Surveys \& Tutorials, IEEE Communication Magazine and IEEE System Journal.

\textbf{Qiang Zheng}
received his B.S. degree from the College of Computer Science and Technology, Shandong University of Technology (SDUT), China, and M.S. degree in telecommunication and information system from Beijing University of Posts and Telecommunications (BUPT), Beijing, China, in 2010, and 2012, respectively. Now, he is currently a Ph.D. candidate in BUPT. His research interests focus on radio resource allocation, performance analysis, and optimization in heterogeneous vehicular networks.

\textbf{Haojun Yang}
received his B.S. degree from Beijing University of Posts and Telecommunications (BUPT), China, in 2014. Since then he has been working toward a Ph.D. degree at BUPT. His research interests include vehicular networks and wireless communications.

\textbf{Lu Hou}
is currently a Ph.D. candidate in the Key Lab of Universal Wireless Communications, Ministry of Education, Beijing University of Posts and Telecommunications (BUPT). He earned his B.Eng. degree from the School of Information and Communication Engineering, BUPT, China, in 2014. Nowadays, he mainly focuses on Soft-define Network (SDN) and resource allocation in mobile cloud computing systems.

\textbf{Long Zhao}  
received the Ph.D. degree from Beijing University of Posts and Telecommunications, Beijing, China, in 2015, where he is currently a lecturer. From April 2014 to March 2015, he was a visiting student at the Department of Electrical Engineering, Columbia University. His research interests include wireless communications and signal processing.

\textbf{Periklis Chatzimisios}
(SM'13) serves as an Associate Professor at the Computing Systems, Security and Networks (CSSN) Research Lab of the Department of Informatics at the Alexander TEI of Thessaloniki (ATEITHE), Greece. Recently he has been a Visiting Academic/Researcher in University of Toronto (Canada) and Massachusetts Institute of Technology (USA). Dr. Chatzimisios is involved in several standardization activities serving as a Member of the Standards Development Board for the IEEE Communication Society (ComSoc) (2010-today) and lately as an active member of the IEEE Research Groups on IoT Communications \& Networking Infrastructure and on Software Defined \& Virtualized Wireless Access. He is the author/editor of 8 books and more than 100 peer-reviewed papers and book chapters on the topics of performance evaluation and standardization activities of mobile/wireless communications, Quality of Service/Quality of Experience and vehicular networking. His published research work has received more than 1500 citations by other researchers. Dr. Chatzimisios received his Ph.D. from Bournemouth University (UK) (2005) and his B.Sc. from Alexander TEI of Thessaloniki (Greece) (2000).

\end{small}

\clearpage

\clearpage

\clearpage

\clearpage

\clearpage

\clearpage

\begin{table}
	\centering
	\caption{Types of autonomous driving control messages in HETVNETs.}
	\begin{tabular}{| c | c | c |}
		\hline
		\textbf{Categories} & \textbf{Message Contents} & \textbf{Examples} \\
		\hline
		\multirow{12}{*}{\makecell{Periodic State \\Messages}} & \multirow{2}{*}{Position} & \multirow{12}{*}{\parbox{7cm}{\emph{Example 1:} When ADVs are traveling on the road, they need to broadcast and report PSMs in an appropriate intervals. \\ \emph{Example 2:} When any ADV is out of the order, it needs to broadcast malfunction messages to warn nearby ADVs to keep distance from it.}} \\
		&  &  \\  \cline{2-2}
		& \multirow{2}{*}{Direction} &   \\
		&  &  \\  \cline{2-2}
		& \multirow{2}{*}{Speed} &   \\
		&  &  \\ \cline{2-2}
		& \multirow{2}{*}{Malfunction} &   \\
		&  &  \\ \cline{2-2}
		& \multirow{2}{*}{Others} &   \\
		&  &  \\
		\hline
		\multirow{10}{*}{\makecell{Action-Triggered \\Messages}} & \multirow{2}{*}{Change Lanes} & \multirow{10}{*}{\parbox{7cm}{\emph{Example 1:} ADV 1 broadcasts ATMs to warn surrounding ADVs before it starts to change lane. \\ \emph{Example 2:} When an emergency vehicle enters into a road segment, it needs to broadcast ATMs to other ADVs to let them pull over.}} \\
		&  &  \\  \cline{2-2}
		& \multirow{2}{*}{Overtake} &   \\
		&  &  \\  \cline{2-2}
		& \multirow{2}{*}{Brake} &   \\
		&  &  \\  \cline{2-2}
		& \multirow{2}{*}{\makecell{Emergency Vehicle \\Avoidance}} &   \\
		&  &  \\  \cline{2-2}
		& \multirow{2}{*}{Others} &   \\
		&  &  \\
		\hline
	\end{tabular}
	\label{table_1}
\end{table}
\clearpage

\end{document}